\documentclass[12pt,a4paper,final]{iopart}
\pdfoutput=1
\usepackage[breaklinks=true,colorlinks=true,linkcolor=blue,urlcolor=blue,citecolor=blue]{hyperref}
\usepackage{latexsym}
\usepackage{graphics}
\usepackage{graphicx}
\usepackage{epsfig}
\usepackage{mathcomp}
\usepackage{amsopn}
\usepackage{amsfonts}
\usepackage{amstext}
\usepackage{amssymb}
\usepackage{amstext}
\usepackage{subfigure}
\usepackage{braket}
\usepackage{iopams}  
\usepackage{wasysym}

\usepackage{epstopdf}
\usepackage{braket}

\begin{document}
\title{Hexagonal Plaquette Spin-{S}pin Interactions and Quantum Magnetism in a Two-dimensional Ion Crystal}
\author{R. Nath$^{1,2,3}$, M. Dalmonte$^{1,2}$, A.~W. Glaetzle$^{1,2}$,  P. Zoller$^{1,2}$, F.~Schmidt-Kaler$^{4}$  and R. Gerritsma$^{4}$}
\address{$^1$Institute for Quantum Optics and Quantum Information of the Austrian Academy of Sciences, A-6020 Innsbruck, Austria}
\address{$^2$Institute for Theoretical Physics, University of Innsbruck, A-6020 Innsbruck, Austria}
\address{$^3$Indian Institute of Science Education and Research, Pune 411 008, India}
\address{$^4$QUANTUM, Institut f\"ur Physik, Johannes Gutenberg-Universit\"at Mainz, D-55099 Mainz, Germany}

\ead{}
\begin{abstract}
We propose a trapped ion scheme en route to realize spin Hamiltonians on a Kagome lattice which, at low energies, are described by emergent $\mathbb{Z}_2$ gauge fields, and support a topological quantum spin liquid ground state. The enabling element in our scheme is the hexagonal plaquette spin-spin interactions in a 2D ion crystal. For this, the phonon-mode spectrum of the crystal is engineered by standing-wave optical potentials or by using Rydberg excited ions, thus generating localized phonon-modes around a hexagon of ions selected out of the entire two-dimensional crystal. These tailored modes can mediate spin-spin interactions between ion-qubits on a hexagonal plaquette when subject to state-dependent optical dipole forces. We discuss how these interactions can be employed to emulate a generalized Balents-Fisher-Girvin model in minimal instances of one and two plaquettes. This model is an archetypical Hamiltonian in which gauge fields are the emergent degrees of freedom on top of the classical ground state manifold. Under realistic situations, we show the emergence of a discrete Gauss's law as well as the dynamics of a deconfined charge excitation on a gauge-invariant background using the two-plaquettes trapped ions spin-system. The proposed scheme in principle allows further scaling in a future trapped ion quantum simulator, and we conclude that our work will pave the way towards the simulation of emergent gauge theories and quantum spin liquids in trapped ion systems.
\end{abstract}

\pacs{37.10.Ty, 37.10.Vz, 03.67.Lx, 05.30.Rt}
\submitto{\NJP}

\maketitle

\tableofcontents

\section{Introduction}


Topological quantum spin liquids are fascinating states of matter supporting topological order and exotic excitations with fractional statistics~\cite{Fmbook}. Recently, there has been intense theoretical activity aimed at underpinning their properties, and identifying microscopic Hamiltonians which could support such states. This activity is mostly focused on the {\it weak} and {\it strong} insulator scenarios~\cite{Fmbook}: in the first case, the focus is on SU(2) invariant spin Hamiltonian on frustrated geometries~\cite{book_frad}, which could potentially be relevant for a series of solid state materials. In the strong insulator scenario, one is instead interested in Hamiltonians displaying exotic constrained dynamics~\cite{Fmbook,nisoli2013,qsl_balents_10}. This second route has been particularly fruitful at the theory level, as the corresponding model Hamiltonians are amenable to powerful analytical techniques and, most importantly, provide a very clean interpretation of quantum spin liquids in a gauge theory framework~\cite{Fmbook,moessner2001b}. However, due to the exotic form of the basic constraints necessary to realize a strong insulator dynamics, many paradigmatic Hamiltonians in this class are still thought of as idealized dynamics without any direct physical counterpart.

Here, we show how a minimal instance of an archetypical example of spin Hamiltonian supporting a topological quantum spin liquid ground state, the Balents-Fisher-Girvin (BFG) model~\cite{balents2002}, can be realized in 2D laboratory trapped ion systems. The key element is to engineer constrained dynamics in a controlled fashion, and in particular, to realize exotic 'plaquette' constraints typical of spin Hamiltonians in the strong insulator scenario. The BFG model is defined on a Kagome lattice, where each site hosts a spin-1/2 degree of freedom, and is
\begin{eqnarray}
\hat H=J_z\left(\sum_{i\in \hexagon}S_z^i\right)^2+\frac{J_{\perp}}{2}\sum_{\langle ij\rangle}\left(S_+^{i}S_-^{j}+h.c.\right).
\label{ham1}
\end{eqnarray}
Here, $S_\alpha^j$ ($\alpha=+,-,z$) are spin operators acting on the site $j$, the first term with $J_z(>0)$ represents Ising interactions around each hexagonal-plaquette, and the second term (with $J_{\perp}$) describes the nearest-neighbour spin exchange interaction (we consider here the variant of the BFG discussed in Ref.~\cite{isakov2012,isakov2011}, where only nearest-neighbor exchange is included). The underlying source of frustration in this model is the emergence of a macroscopic classical ground-state degeneracy due to the local hexagonal plaquette constraints imposed by the first term in the Hamiltonian, i.e., $\sum_{i\in\hexagon}S_z^i=0$. Hence, the ground state configurations are such that each hexagon in a Kagome lattice consists of three up spins and three down spins. This is reminiscent of the ice rule in spin-ice models~\cite{nisoli2013,qsl_balents_10} in which, close to each vertex of a square lattice, two spins point inwards, and two spins outward. Once quantum fluctuations are turned on, $J_\perp \ll J_z$, the system is effectively described by a quantum dimer model on the dual (triangular) lattice~\cite{moessner2001b,moessner2001}. Thus, in contrast to the 2D version of spin Ice, whose dynamics is described by a compact U(1) gauge theory, the low-energy physics of the BFG model is different, as the underlying lattice on top of which the quantum dimer model is defined is not bipartite. In those cases, the emergent degrees of freedom are $\mathbb{Z}_2$ dynamical gauge fields~\cite{Fmbook}, which undergo deconfinement and stabilize a topological quantum spin liquid state~\cite{balents2002,isakov2011,sheng2005}.

Our proposal to implement the BFG model is based on the remarkable achievements in trapped ion technology in the last decades~\cite{schn-rvw12,qs_bla_12}. These highly controllable quantum systems have led to a variety of proposals and realizations of quantum simulators, such as the topological hexagonal Kitaev model \cite{leib-kitaev_11}, fermionic lattices \cite{casa_ferm_12}, SU(2) Ising models \cite{mg_qs_08,kim_tm_09,yb_edward_10,Lanyon_11,range_britton_12} -including frustrated magnetism \cite{yb_kim_10,islam_frus_13,berm_fr_12} and, very recently, to the observation of entanglement dynamics in spin chains \cite{entang_rich_14,entang_roos_14}. Besides these studies, which are implemented in linear-(1D) ion crystals, quantum simulation of various spin models has also been proposed in 2D ion-crystals \cite{porras_ss_04,porras_sm_06,Bermudez_11} and various experimental approaches are studied, such as 2D Paul traps \cite{Landa_12a,Landa_12b,Kaufman_12,2D_yoshi_14}, Penning traps \cite{penn_wine_98,penn_baugh_98}, micro- \cite{arc_zoll_00} and multi zone- \cite{arc_wine_02,arc_mon_13} trap arrays. Recent work in a Penning trap has demonstrated controllable spin-spin interactions between a few hundred ionic spins~\cite{range_britton_12}, whereas studies in Paul traps show excellent prospects for implementing such interactions as well~\cite{Kaufman_12}.


Here, we show how to generate hexagonal-plaquette spin-spin interactions by (anti) pinning appropriate ions in a 2D ion crystal by external potentials which we describe below. The spin-spin interactions are mediated by a spin dependent optical dipole force interacting with a localised phonon mode that appears as a result of this pinning. This localised phonon mode involves all ions in the plaquette but rapidly decays outside it. The resulting spin-spin interactions on the hexagonal-plaquette represent a key building block for quantum simulators aimed at studying the fundamental nature of frustrated quantum magnetism with emergent gauge fields \cite{book_frad,qsl_balents_10,qsl_moes_11} and may open up new directions in ion-based quantum simulators. By pinning multiple ions, more individual plaquette interactions can be engineered thus allowing for scaling up the quantum simulator. In the present paper, we discuss the minimal instance of a two-plaquette implementation, which should lie within current experimental reach. Note that our current studies also fit well into the recent trends of cold atom-investigations on synthetic gauge fields \cite{gauge_rvw_11}, in particular on dynamical gauge fields \cite{dyga_lew_13,dyga_tew_06,dyga_ban_13,dyga_cir_13,dgf_alex_14-1,dgf_alex_14-2,wiese2013}. 

The ion pinning can be accomplished by a standing wave light field focused onto the ion to be pinned. Such optical potentials can reach curvatures corresponding to trap-frequencies in the MHz range. The proposed setup resembles those encountered in recent studies involving ions interacting with strong light-fields, such as in anomalous diffusion of an ion \cite{diff_ion_97}, preparing nonclassical motional states \cite{ncs_cir_94}, and, in particular, modifying the trapping potential locally in a rf-trapped ion crystal for quantum simulations \cite{qs_cir_2008,qs_haff_11}. Additionally, optical trapping of an ion has been shown experimentally using either a single beam dipole trap \cite{ot_schtz_10,ot_schtz_12,ot_schtz_14} or an optical lattice \cite{ol_ion_sc_12,ol_ion_dre_12,ol_ion_vule_13,ol_ion_sg_14}. An interesting alternative option would be to dress the ion of interest with a Rydberg state, such that the dipole moment induced by the ionic trapping field causes a change in local trapping frequency for the ion~\cite{Muller_NJP_09,FSK_NJP_11}. Here, no spatial variation of the pinning laser would be required and additional microwave fields could be employed for fine-tuning~\cite{Li_PRL_12,Li_PRA_13}.

The paper is organized as follows: in section \ref{setup} we discuss the physical setup of the ion-laser system and the governing atom-light interaction Hamiltonian. The spin models are derived from the interaction Hamiltonians, we show how to generate  spin-spin hexagonal interactions in a single plaquette using $N=7$ and two plaquettes using $N=19$ ion crystals. In section \ref{mag} we discuss the ground state properties and magnetization dynamics for the single and double plaquettes spin-systems, and the results are compared with that of the original BFG model. Realistic numbers for the laser parameters for a calcium ion-setup are given in section \ref{ca-ion}. Finally, we summarize the paper with an outlook in section \ref{summ}.


\section{The Balents-Fisher-Girvin model in a trapped ion quantum simulator: building blocks}
\label{setup}
\subsection{Spin-spin interactions in a two-dimensional ion crystal}
\label{model}
\begin{figure}[hbt]
\centering
\includegraphics[width= 1.0\columnwidth]{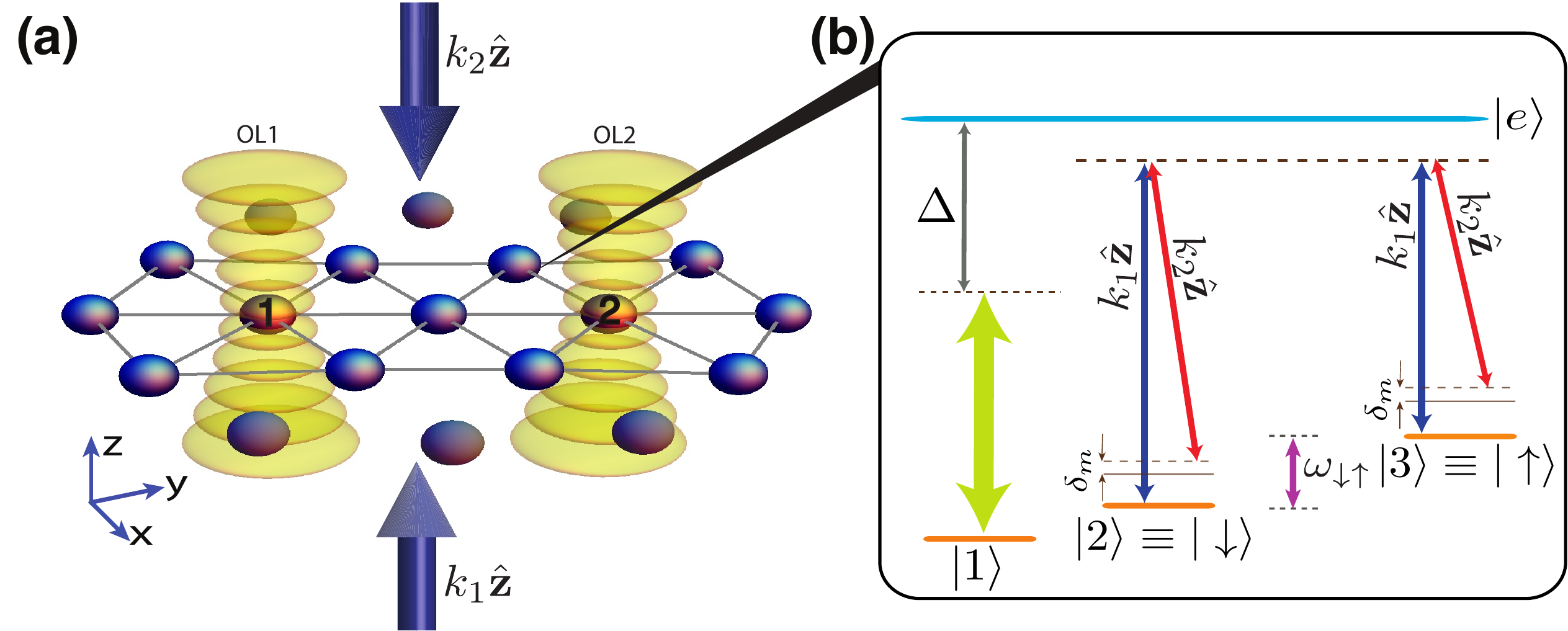}
\caption{\small{(a) The schematic picture of the ion-laser system for generating hexagonal-plaquette interactions. A pair of counter-propagating Raman fields with wave vectors (frequencies) $k_1 \hat{\bf z}$ ($\omega_1$) and $k_2 \hat{\bf z}$ ($\omega_2$) are used for inducing spin-spin couplings. OL1 and OL2 (yellow in colour) are two optical lattices focused on ions 1 and 2 to modify their transversal (along the $z$ axis) trapping frequencies locally. (b) The level scheme: the internal level structure consists of three long-lived low-lying states and a manifold of excited states. The Raman lasers (blue and red arrows), detuned by $\delta_m$ with respect to the frequency of a motional mode, generate state dependent optical shifts in the spin states $|2\rangle\equiv|\downarrow\rangle$ and $|3\rangle\equiv|\uparrow\rangle$ with an energy separation of $\hbar\omega_{\downarrow\uparrow}$, which simulate the spin-spin interactions of the type $S_z^i\otimes S_z^j$ between different ions $i$ and $j$.  The ions 1 and 2  prepared in state $|1\rangle$ experience an additional optical lattice that is created by very far detuned laser fields (yellow arrow) from the excited state $|e\rangle$ with a detuning $\Delta_{OL}$. The lattices modify their local trapping frequencies along the $z$ axis. Additional laser fields, used for generating quantum dynamics in the spin system, are not shown in the figure for the sake of clarity.}}
\label{fig:setup} 
\end{figure}

A sketch of our proposed experiment is shown in figure \ref{fig:setup}. The system consists of a self-assembled 2D ion crystal confined in the $xy$ plane by what we, for simplicity, assume to be a radially symmetric trap  with trap-frequencies: $\omega_{x,y}\ll\omega_{z}$. The internal level structure of the ions consists of three long-lived low-lying atomic states, these can be either three Zeeman states in the ground-state manifold or a combination of metastable and ground states.  The states  $|2\rangle\equiv\mid\downarrow\rangle$ and $|3\rangle\equiv\mid\uparrow\rangle$ encode the spin-half states of the BFG model. The ion crystal supports quantized phonon modes that are used to transmit spin-spin interactions between the ions via a Raman pair of counter-propagating laser fields as shown in figure~\ref{fig:setup}b. Ions sitting in the middle of an hexagonal sublattice (see, e.g., ions 1 and 2 in Fig. 1) are prepared in the state $|1\rangle$ and strongly pinned by a standing wave laser field that is far detuned from any excited states $\{|e\rangle\}$ to locally modify the phonon spectrum.

The details of laser mediated spin-spin interactions in trapped ion systems have been extensively described elsewhere~\cite{schn-rvw12} and are briefly discussed in \ref{ss-int}. Two counter propagating laser fields are far detuned  from the transitions $\mid\downarrow\rangle \leftrightarrow |e\rangle$ and $\mid\uparrow\rangle \leftrightarrow |e\rangle$. When the Rabi frequencies  of these lasers are much smaller than the detunings, the excited state $|e\rangle$ can be adiabatically eliminated. To engineer spin-spin interactions of the form $S_z \otimes S_z$, the frequency difference of the laser beams $\omega_I$ is tuned close to the phonon frequencies $\omega_m$, where $m$ denotes the particular phonon mode and $\delta_m=\omega_I-\omega_m$. Throughout the paper we assume to be in the Lamb-Dicke regime, $\eta_m^i\sqrt{\langle n_m+1 \rangle} \ll 1$, where $\eta_m^i=q_{m0}b_m^ik_I$ are the Lamb-Dicke parameters, $n_m$ the number of phonons in mode $m$, $q_{m0}=\sqrt{\hbar/2M\omega_m}$ with $M$ the mass of an ion and ${\bf b}_m=\{b_m^i\}$ are the (normalised) phonon eigenvectors. As long as $\eta_m\Omega_I(n) \ll \delta_m$, with the two photon Rabi frequency $\Omega_I(n)=\Omega_1(n)\Omega_2(n)/\Delta$ for each laser and state $n=\uparrow, \downarrow$ and $\Delta_1 \approx \Delta_2 \gg \Omega_{1,2}(n)$ the detuning for each laser, the Raman laser only transiently excites phonons and we can integrate out the phonon degrees of freedom altogether to obtain effective spin-spin interactions between ion $i$ and $j$ of the form $\hat H_{ZZ}=\sum_{i,j}J_z^{ij}S_z^i\otimes S_z^j$ with coupling matrix

 \begin{equation}
J_z^{ij}=\sum_{m=1}^{N}\frac{4\Omega_I^i\Omega_I^j  \eta_m^i\eta_m^j}{\delta_m},
\label{ss-j}
\end{equation}
\noindent with $\Omega_I^i=\Omega_I^i(\downarrow)-\Omega_I^i(\uparrow)$. The strength and range of $J_z^{ij}$ are determined by the Rabi frequencies $\Omega_I^i$, the detuning $\delta_m$ and the amplitude of oscillation of each ion in the $m^{\rm {th}}$ mode through the Lamb-Dicke parameters $\eta_m^i$. Note that we only assumed to be in the Lamb-Dicke regime in deriving the spin-spin interaction Hamiltonian, so ground state cooling is not necessary~\cite{qc_mol_99}.

We can also obtain effective spin-spin interactions in equation \ref{ss-j} for the case when $\eta_m\Omega_I \ll \delta_m$ does not hold, but in this scenario, the phonons in each mode $m$ only return to their initial state at particular times $1/\delta_m$. A favorable situation occurs when $\eta_m\Omega_I \ll \delta_m$ for $m \neq n$, that is for all phonon modes except one. In this case, the spin-spin interaction takes the form of equation \ref{ss-j} at times~$t_{gate}=k/\delta_n$ with $k$ an integer. This situation is usually employed when considering quantum gate operations~\cite{qc_mol_99}. Note that any remaining spin-motion entanglement causes decoherence in the spin state and appears as an error in the quantum simulator.

Using additional sets (two) of Raman fields we can also generate spin-spin couplings $J^{ij}_{\perp}$ of the form $\hat H_{XX}+\hat H_{YY}\sim S_x\otimes S_x+S_y\otimes S_y$~\cite{kim_tm_09,range_ma_14}. For that we assume $\omega_I$ is tuned close to $\omega_{\uparrow\downarrow}\pm\omega_m$, a situation in which we have spin flip as well as the creation of phonons. The expression for $J_{\perp}^{ij}$ is still given by equation \ref{ss-j}, but with a re-defined detuning $\delta_m=\omega_I-(\omega_{\uparrow\downarrow}\pm\omega_m)$. Note that the different spin-spin couplings, $J^{ij}_z$ and $J^{ij}_{\perp}$, can be independently controlled by the laser  parameters of the corresponding Raman fields. The corrections to the total Hamiltonian,
\begin{equation}
\hat H= \sum_{i<j}J_z^{ij}S_z^i\otimes S_z^j+\sum_{i<j}J_{\perp}^{ij}\left(S_x^i\otimes S_x^j+S_y^i\otimes S_y^j\right),
\end{equation} 
arising from the non-commutativity of the different coupling Hamiltonians are oscillatory and can be neglected \cite{porras_ss_04,tobias_ss_13,deng_ss_05}. 
As experimental complexity increases when considering more laser beams, in \ref{hxy} we also discuss a reduced Hamiltonian of the form $\hat H_{XZ}=\hat H_{XX}+\hat H_{ZZ}$, in which the couplings $\hat H_{YY}$ are not included. Notice that, within the perturbation theory on the gauge-invariant manifold, the strong coupling Hamiltonians of the two different cases are very similar at lowest orders. We found that, in the $\hat H_{XZ}$ case, the underlying physics is almost unaffected when compared to the full setup. Alternatively, the simulation may be implemented by a stroboscopic sequence of "Trotter steps" in which each term in the Hamiltonian is switched on for a short time $\Delta t$ consequetively~\cite{Lanyon_11}. In this concatenated form, errors due to non-commutativity between the terms in the Hamiltonian scale with the discretized stepsize squared: $O(\Delta t^2)$. This approach also has the advantage that less independent laser beams are necessary.

\subsection{Single hexagonal plaquette in an $N=7$ ion crystal}
\label{sec_single_p}

From equation (\ref{ss-j}) it is clear that phonon-mode spectrum is of central importance to the spatial form of the spin-spin interactions, as it determine the Lamb-Dicke parameters $\eta_m^j$ as well as the mode detuning $\delta_m$. In single species ion crystals, in which all ions experience the same trapping frequency, the phonon modes are highly collective, which results in long range spin-spin interactions. Here, we discuss how to generate anti-ferromagnetic spin-spin interactions displaying a hexagonal-plaquette pattern in a 2D ion crystal. By plaquette pattern we imply that each of the six spins occupying the corners of a hexagon interact with every other spin in the same hexagon with the same strength irrespective of their inter-spin separation. First, we discuss how to generate it in a minimal setup - a crystal consisting of seven ions - and then extend to a larger crystal made of 19 ions.

\begin{figure}[hbt]
\centering
\includegraphics[width= 1.0\columnwidth]{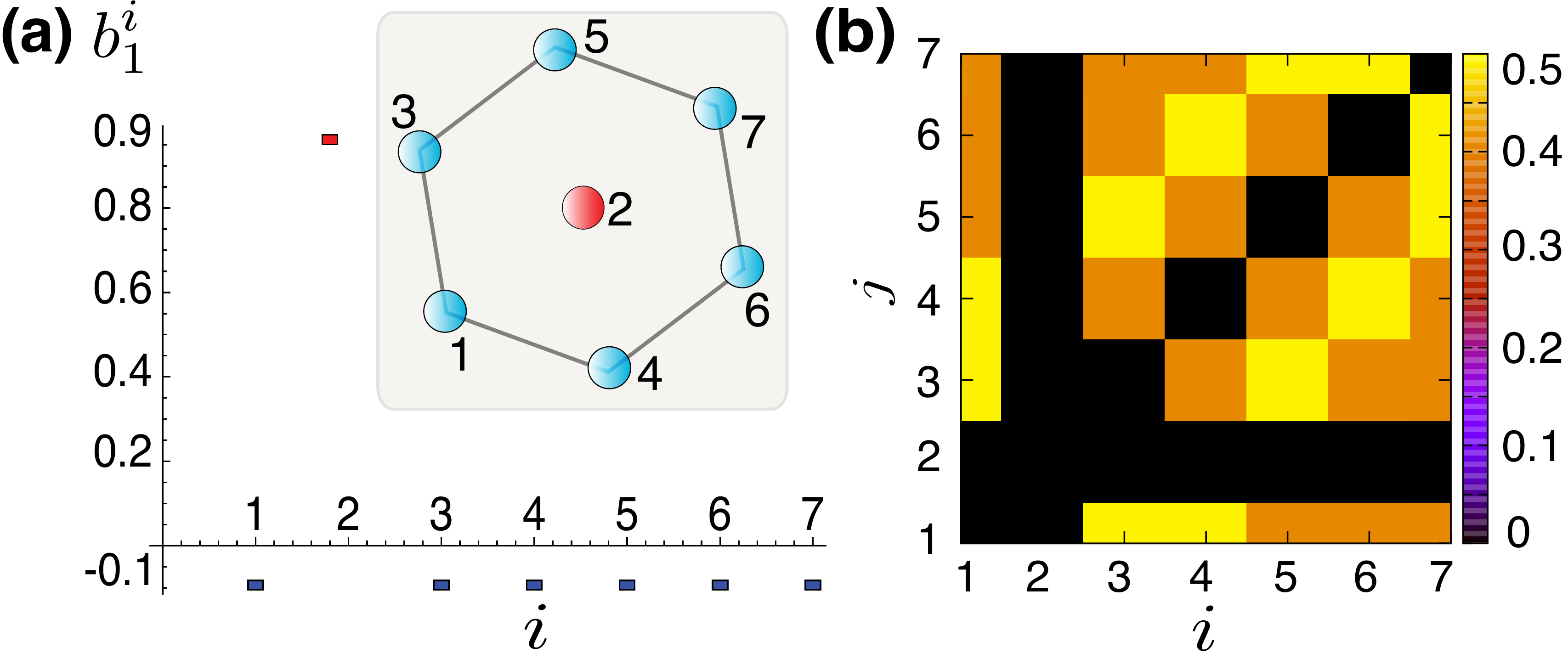}
\caption{\small{(a) The inset shows the equilibirum position of 7 ions in a radially symmetric trap with trapping frequencies: $\omega_{x,y}=2\pi\times 1$ MHz and $\omega_z=2\pi\times 3$ MHz. Since the central ion experiences an additional pinning lattice, the effective trapping frequency of it along the $z$ axis is reduced to $\tilde\omega_z=2\pi\times 2.7$ MHz.   The eigenvector ($b_1^i$) of the lowest transversal mode  of the ion crystal is shown in figure \ref{fig:eve7}a. It accounts mostly the oscillation of the central ion, and the ions in the outer hexagonal ring oscillates with same amplitude and are in-phase. (b) The dimensionless spin-spin couplings $J_{ij}/\omega_0$, where $\omega_0=\left(\hbar k_I^2/8M\right)\left(\omega_x/\Omega_I\right)^2$, with $\delta_{1}=2\pi\times 10$ kHz, the detuning from the lowest TM. The slight imperfections from the BFG plaquette interactions arise from the off-resonant coupling to higher TMs. For sufficiently small value of $\tilde\omega_z$, the lowest TM mode of the 2D crystal become energetically unstable, i.e. $\nu_1=0$ and leads to a structural phase transition in which the 2D character of the whole crystal has been lost \cite{2D_yoshi_14}. For the set of parameters in figure \ref{fig:eve7} it happens when $\tilde\omega_z<\omega_z/2\sim 2\pi \times 1.5$ MHz, which restricts the frequency due to the optical lattice potential to values $\omega_{OL}/\omega_z < \sqrt{3/4}$.}}
\label{fig:eve7} 
\end{figure}

The equilibrium configuration of an ion-crystal with 7 ions in a radially symmetric trap is shown in the inset of figure \ref{fig:eve7}a. Although they form a triangular lattice, we picture it as a hexagonal structure with an ion in its centre. The central ion is prepared in the atomic state $|1\rangle$, (see figure \ref{fig:setup}), which is immune to the Raman fields used for spin operations, but experiences an additional one-dimensional pinning optical lattice applied normal to the plane of the crystal, i.e. along the $z$-axis.  The rest of the ions occupying the spin states form a hexagon shape. The pinning laser modifies the transversal trapping frequency of the central ion to $\tilde \omega_z=\sqrt{\omega_z^2-\omega_{OL}^2}$ compared to the other ions, when the maximum of the lattice potential is focused at its equilibrium position, where $\omega_{OL}$ is the harmonic-frequency of the local potential at the ion position due to the optical lattice. For a red detuned lattice laser and the ion trapped in an antinode, the pinning lattice relaxes the confinement of the central ion along the transversal direction and we see below how it affects the phonon spectrum. In the following, we restrict ourselves to pinning-lattice parameters such that the equilibrium positions of the ions are not altered once it is switched on.

 \begin{figure}[hbt]
\vspace{0.cm}
\centering
\includegraphics[width= 1.\columnwidth]{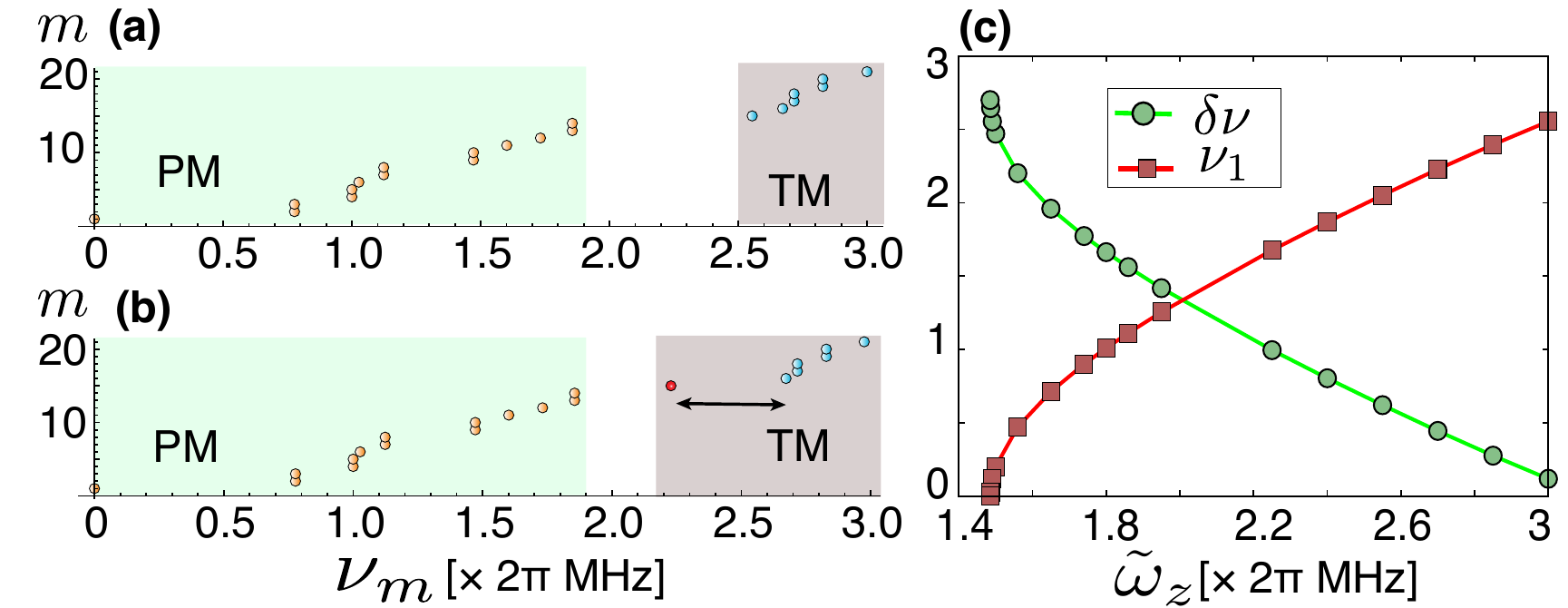}
\caption{\small{(a) The normal mode spectrum of a trapped 2D ion crystal made of 7 ions,  in a radially symmetric trap of  frequencies $\omega_{x,y}=2\pi\times 1$ MHz and $\omega_z=2\pi\times 3$ MHz. The orange-filled bubbles are the in-plane modes (PM) and the blue-filled ones are the transversal modes (TM). (b) The same in the presence of the pinning optical lattice.  The optical lattice reduced the trapping frequency of the central ion along the $z$ axis to a  frequency $\tilde \omega_z=2\pi\times 2.7$ MHz. The red-filled bubble indicates the lowest TM and the frequency of it is shifted to a lower value due to the presence of pinning field (shown by a black arrow). (c) The lowest transversal mode $\nu_1$ (filled squares) and the energy difference between the lowest two transversal modes $\delta\nu=\nu_2-\nu_1$ (filled circles) as a function of $\tilde\omega_z$ are shown. For $\tilde\omega_z<\omega_z/2$ the mode $\nu_1$ gets energetically unstable and the 2D nature of the crystal is lost. }}
\label{fig:ev7} 
\end{figure}

 The normal-mode spectrum consists of 14 $xy$ (in-plane) and 7 $z$-(transversal) phonon modes, see figures \ref{fig:ev7}a and \ref{fig:ev7}b. Due to the 2D character of the crystal, the in-plane (PM) and the transverse modes (TM) are completely de-coupled. The TMs account for the oscillation of ions along the tightly confined direction, and lie on top of the energy spectrum, see figure \ref{fig:ev7}a. The pinning affects mostly the TMs in two ways: (i) it shifts the eigen values and (ii) modifies the mode-structure ${\bf b}^i_m$. In particular, the frequency $\nu_1$ of the lowest TM (see figure \ref{fig:eve7}a), which accounts mostly for the oscillation of the central ion, moves down to lower values as $\tilde\omega_z$ decreases, see figure \ref{fig:ev7}c. 

The eigen vector of the lowest TM in the presence of pinning field is shown in figure \ref{fig:eve7}a. Note that it exhibits a hexagonal character in which all the ions in the hexagonal ring oscillate with the same amplitude and in phase. When the Raman beat frequency is tuned close to this mode, the resulting spin-spin couplings $J_z^{ij}$ exhibit a hexagonal-plaquette pattern as shown in figure \ref{fig:eve7}b. There are slight imperfections from the ideal hexagonal pattern discussed in the BFG model (equation \ref{ham1}) due to the off-resonant couplings between the Raman fields and other TMs. These imperfections can be reduced by providing a sufficient gap between the first TM and the rest. As shown in figure \ref{fig:ev7}c, this can be done by tuning the frequency of the local harmonic potential due to the pinning optical lattice at the central ion position. 

If one considers a Paul trap,  the anharmonic coupling between the in-plane micro-motion and the transverse modes has to be taken into account, which, as recently shown \cite{paul_duan_14}, only introduces an overall renormalization in the ion positions without effecting the normal mode structure. While an overall shift in the transversal mode frequencies should also to be taken into account, but it has no effect on our underlying scheme for generating spin-spin interactions. Hence, we can safely neglect those effects in our calculations. In the spectrum, there exists a zero-energy in-plane mode due to the rotational symmetry of the ion crystal, corresponding to the free rotation of ion crystal in the $xy$ plane. In real experiments, any weak stray fields will break this symmtery and the crystal will be trapped in a stable configuration. 
  

\subsection{Double hexagonal plaquette in an $N=19$ ion crystal}
\label{ion19}

Having established the basic idea of hexagonal plaquette interactions via  an ion crystal of seven ions, we expand our discussion to two plaquettes in a larger crystal. Here, we consider a crystal of 19 ions in a radially symmetric trap. In order to create two plaquettes, we require two pinning standing wave lasers as shown in figure \ref{fig:setup}a and focus them such that the maxima of lattice potentials lie at the equilibrium positions of the two ions populated in state $|1\rangle$. These two ions, as in the previous case, experiences a shallower trap along the $z$ axis with respect to the other ions.

\begin{figure}[hbt]
\centering
\includegraphics[width= 0.9\columnwidth]{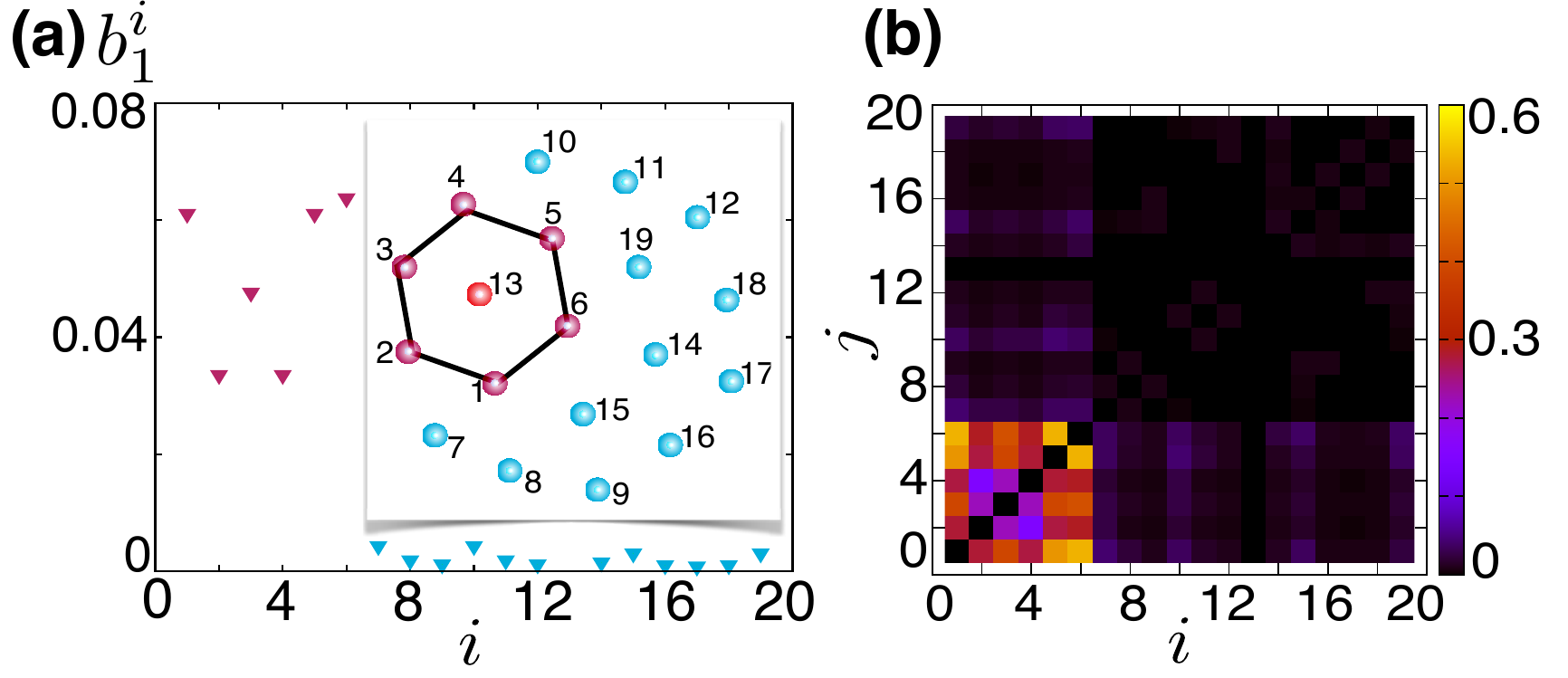}
\caption{\small{(a) The eigen vector for the lowest transversal mode in a 19-ion crystal in the presence of a pinning lattice on ion 13. The inset shows the pinning scheme in which the ion in the centre of the hexagon is populated in the state $|1\rangle$, which experiences an additional optical lattice. The mode accounts mostly the oscillation of the central ion (not shown in the plot) and the ions in the hexagonal ring. Note that ions outside the hexagon hardly oscillate. (b) The dimnensionless spin-spin couplings:  $J_z^{ij}/\omega_0$, where $\omega_0=\left(\hbar k_I^2/8M\right)\left(\omega_x/\Omega_I\right)^2$, for $\delta= 2\pi\times10$ kHz the detuning from the lowest transversal mode. The trapping frequencies are $\omega_x=\omega_y=2\pi\times1$ MHz and $\omega_z=2\pi\times3.5$ MHz and the pinned ion experiences an effective shallow trap along the $z$ axis with a frequency of $\tilde\omega_z^1=2\pi\times2.1$ MHz. Note that the interaction pattern is well localized in the hexagonal ring.}}
\label{fig:ss19-1} 
\end{figure}

The parameters of the pinning lattices are chosen such that the equilibrium positions are not altered. The trapping parameters are taken to be $\omega_x=\omega_y= 2\pi\times 1$ MHz and $\omega_z=2\pi\times 3.5$ MHz and the effective harmonic frequency at the pinned ions along the $z$ axis are $\tilde \omega_z^{(1)}=2\pi\times 2.1$ MHz and $\tilde \omega_z^{(2)}=2\pi\times 2.45$ MHz respectively for ion 1 and ion 2. The reason for the asymmetry in the pinning lattices parameters is discussed below. With these parameters, the two lowest TMs have the plaquette character and they comprise mostly of the oscillations of the two pinned ions. If we switch off either one of the pinning lattices, we retrieve the same physics discussed in the case for one plaquette. As an example, we switch off the pinning lattice on ion 2 and the ion 1 experiences an effective harmonic trapping frequency of $\tilde\omega_z^1=2\pi\times 2.1$ MHz. The resulting lowest TM is shown in figure \ref{fig:ss19-1}a. It accounts mostly the oscillation of ion 1 (not shown in the figure), and is well localized among the six ions surrounding it, forming a hexagonal oscillation pattern. 

The resulting spin-spin interactions, when the Raman fields are detuned close to this mode, exhibit the hexagonal-plaquette pattern (see figure \ref{fig:ss19-1}b) similar to what we have discussed in section \ref{sec_single_p} for the case of 7-ion crystal. This point constitutes one of the main results of our paper: even in a larger crystal, by engineering phonon modes we can create localized spin-spin interactions, which in our case exhibit a hexagonal pattern. Compared to the smaller crystal, here the imperfections are slightly higher. This can be understood from the structure of the plaquette eigenmode itself: the amplitude of oscillation is not the same for all the six ions in the hexagon (see figure \ref{fig:ss19-1}a), which is contrary to the $N=7$ case. This arises due to the asymmetry in the number of nearest-neighbor ions for each ion in the hexagonal ring. A second reason is that the number of modes increases linearly with the system size, but in our case the coupling to those modes are well suppressed by the large detunings $\delta_m$ from the corresponding $m^{\rm th}$ mode. Note that if we pin ion 2 instead of ion 1, we create a hexagonal plaquette around the second ion.
 
\begin{figure}[hbt]
\centering
\includegraphics[width= 0.9\columnwidth]{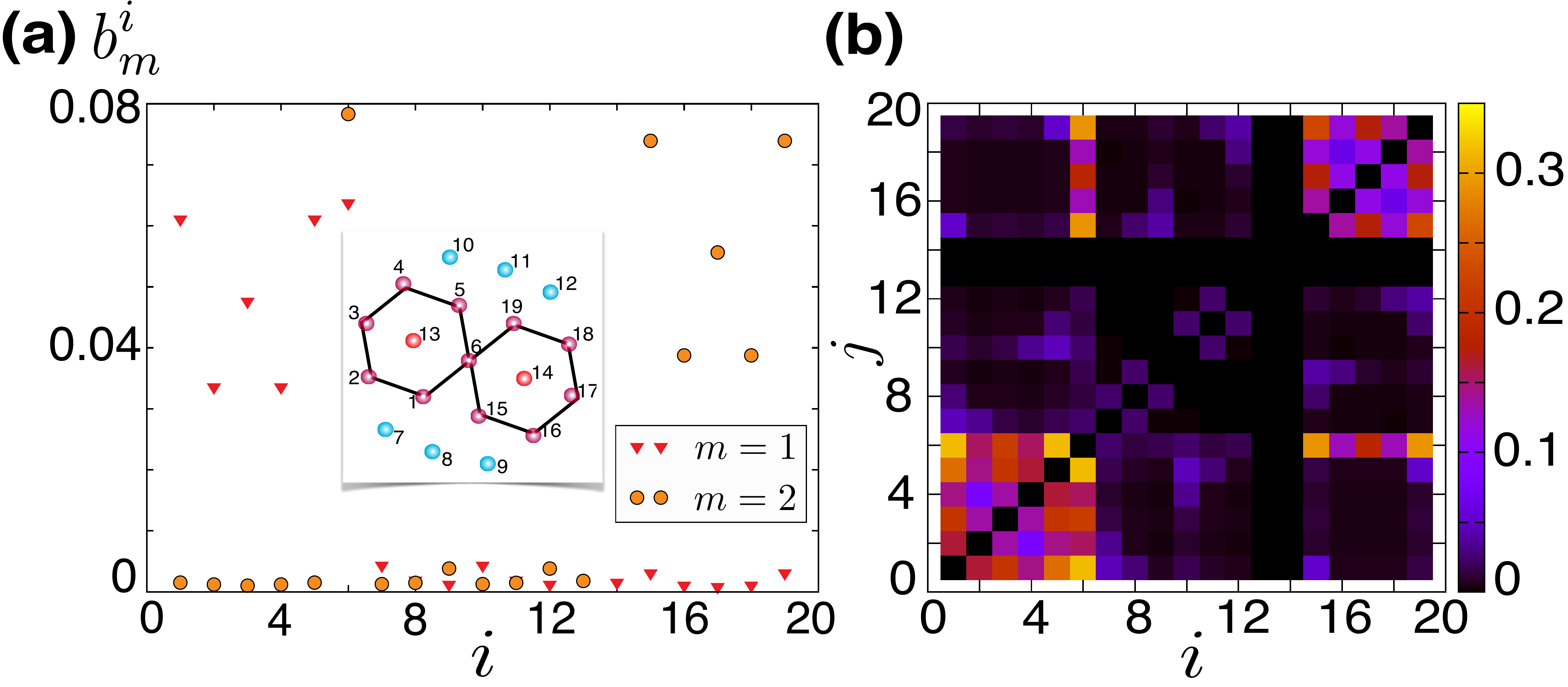}
\caption{\small{(a) The eigen vectors of the two lowest transversal modes in a trapped 19-ion crystal with  trapping frequencies: $\omega_x=\omega_y=2\pi\times1$ MHz and $\omega_z=2\pi\times3.5$ MHz, in the presence of  pinning lattices on ions 13 and 14. The pinning lattices reduce the effective trapping frequencies of ions 13  and 14, along the $z$ axis to $\tilde\omega_z^1=2.1$ MHz and $\tilde\omega_z^2=2.45$ MHz respectively. The inset shows the equilibrium configuration of the corresponding ion crystal. The two localized modes: $b_1^i$ and $b_2^i$ account mostly the oscillation of ion 13 and 14 respectively (not shown in the plot). (b) The dimensionless spin-spin couplings:  $J_z^{ij}/\omega_0$ with $\omega_0=\left(\hbar k_I^2/8M\right)\left(\omega_x/\Omega_I^1\right)^2$, for  $\Omega_I^2/\Omega_I^1=1.1$ and $\delta=2\pi\times 20$kHz. Note that ion 6 is shared by both the plaquettes.}}
\label{fig:ss19-2} 
\end{figure}

 {\em Two plaquettes case}: To create two plaquettes, we switch on both pinning lattices. In this case, the {\it two} lowest phonon modes are used to engineer the plaquette interactions. If we have identical pinning lattices, the two modes become degenerate. Then, dressing them with a single pair of Raman fields may generate an arbitrary superposition of phonon states: $c_1{\bf b}_1\pm c_2{\bf b}_2$, where ${\bf b}_{1,2}$  are the eigen vectors of the two lowest TMs and $|c_{1,2}|^2$ provide us the population of the respective phonon modes. The resulting spin-spin interactions do not possess the plaquette character. Hence, we introduce an asymmetry between the parameters of the pinning lattices such that the degeneracy is lifted as well as the modes are well separated in the spectrum. Then, we use two different pairs of Raman fields to induce spin-spin couplings, in which one is detuned near the mode ${\bf b}_1$ and the second one is near the mode ${\bf b}_2$ with the same detuning $\delta$. Locally, we can control the interaction strengths in each plaquette by tuning the effective coupling strengths due to two Raman fields, i.e., $\Omega_I^1$ and $\Omega_I^2$; in the particular example shown in figure \ref{fig:ss19-2}b we choose $\Omega_I^2/\Omega_I^1=1.1$ to obtain the same interaction strengths in both plaquettes. The equilibrium configuration of the 2D ion crystal and the eigenvectors of the modes ${\bf b}_1$ and ${\bf b}_2$ are shown in figure \ref{fig:ss19-2}a. In figure \ref{fig:ss19-2}a, the ions labelled 13 and 14 are prepared in the atomic state $|1\rangle$ and pinned by the optical lattices. The resulting spin-spin couplings $J_z^{ij}$ shown in figure \ref{fig:ss19-2}b are well localized in each hexagonal plaquette and crucially inter-plaquette interactions are negligibly small. $J_z^{ij}$ constitutes the Ising part of the Hamiltonian in equation \ref{ham1}.
 
\begin{figure}[hbt]
\centering
\includegraphics[width= 0.9\columnwidth]{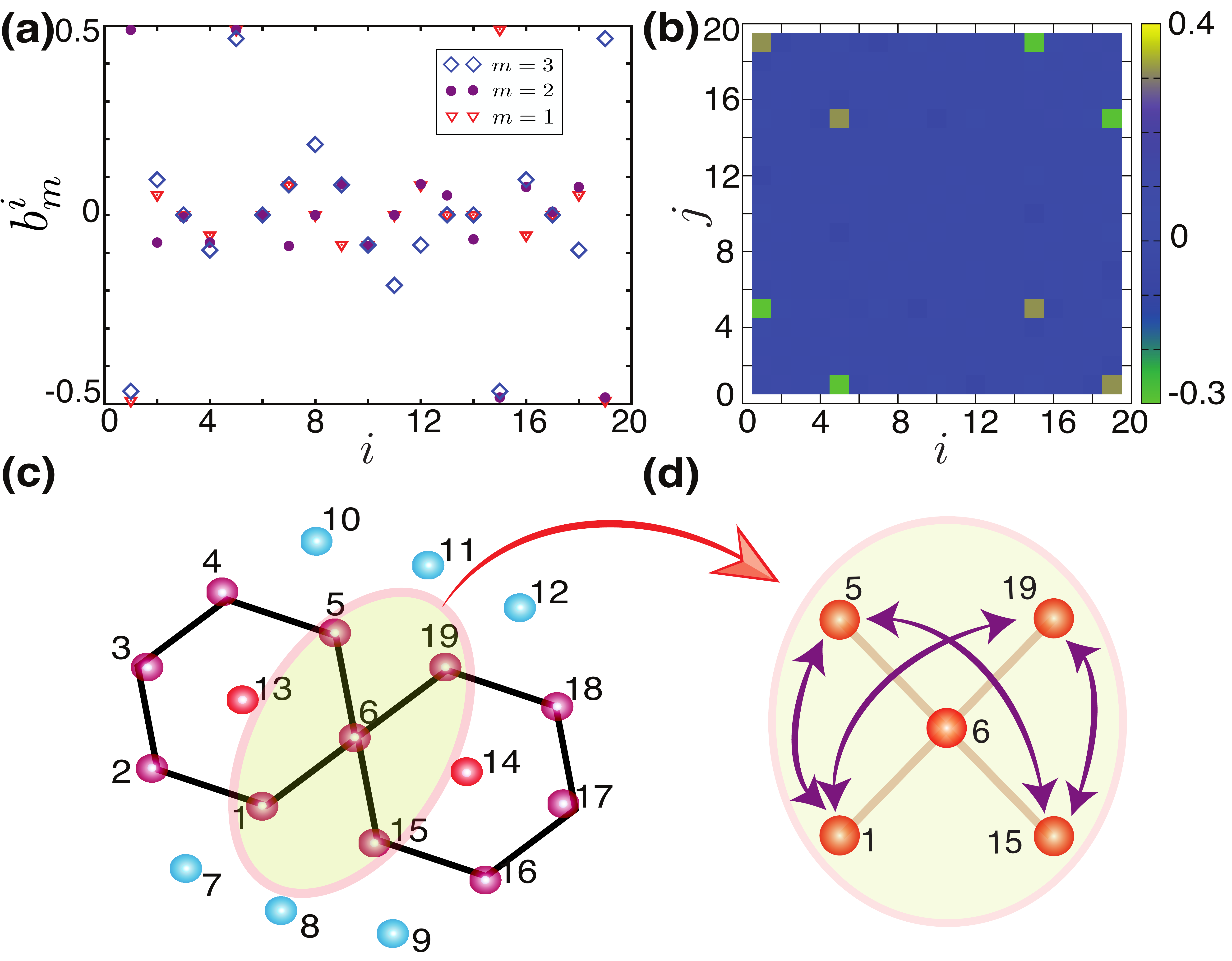}
\caption{\small{(a) The eigen-vectors of three modes lying above the plaquette modes, which are used for designing the hopping dynamics, in a trapped $N=19$ ion crystal with  trapping frequencies: $\omega_x=\omega_y=2\pi\times1$ MHz and $\omega_z=2\pi\times3.5$ MHz, in the presence of  pinning lattices on ions 13 and 14. The pinning lattices reduce the effective trapping frequencies of ions 13  and 14, along the $z$ axis to $\tilde\omega_z^1=2.1$ MHz and $\tilde\omega_z^2=2.45$ MHz respectively. (b) The dimensionless (hopping) spin-spin couplings:  $J_{\perp}^{ij}$ for $\delta_{\perp}=-2\pi\times 60$kHz  w.r.t. the mode $m=1$ shown in (a). (c) shows the ion crystal and (d) shows the different hopping channels created by the $J_{\perp}^{ij}$ in figure \ref{fig:jhop}b.}}
\label{fig:jhop} 
\end{figure}

{\em The hopping parameter $J_{\perp}$} is generated by additional Raman fields, again mediated through transversal modes. The nature of $J_{\perp}$ is also vital in determining the ground state properties as well as the magnetization dynamics of the emulating spin Hamiltonian. As short-range hopping is preferable, e.g. of nearest-neighbor type~\cite{isakov2011}, we identify three lowest TMs in the spectrum, labelled as $m=$1, 2 and 3, see figure \ref{fig:jhop}a. They lie just above the plaquette modes and are also well separated from them, guaranteeing that the Ising part is not influenced when addressing with the Raman fields. Note that these modes are shared among the two corner-sharing triangles at the centre of the crystal, and hence the resulting hopping dynamics may introduce very interesting inter-plaquette spin dynamics, for example {\em ring exchange} among the four ions $\{1,5,15,19\}$ as shown schematically in figure \ref{fig:jhop}d, results in charge hopping between the two plaquettes, (see section \ref{mag} for the consequent magnetization dynamics). The fields are detuned by $\delta_\perp=-2\pi \times 60$ kHz from the $m=1$ mode. The calculated couplings are shown in figure \ref{fig:jhop}b and the different hopping channels are schematically shown in figure \ref{fig:jhop}d.  
This completes our modeling of the microscopic Hamiltonian in an ion crystal for emulating the BFG model in equation \ref{ham1} in a small setup consists of two hexagons. The strategy is scalable to larger systems owing to its analogue nature.

\section{Many-body Physics and Quantum Magnetism in a double-plaquette quantum simulator}
\label{mag}

In this section we discuss the many-body physics associated with the above mentioned spin systems with hexagonal-plaquette interactions. In particular, we focus on the ground state properties and magnetization dynamics. First, we consider the plaquette interactions alone, and address the role of the imperfections on the interaction patterns at the classical level. Then, we show how gauge-invariance, embodied by the magnetization conservation on each plaquette, is affected by the introduction of quantum fluctuations, and compare the realistic scenario with the ideal BFG model. Finally, we show how interesting many-body dynamics can be observed in a minimal system involving two plaquettes, within reach of state-of-the-art experiments with trapped ion crystals, by exploring the dynamics of a single charge on top of the gauge-invariant background.

\subsection{Ground state properties}

\begin{figure}[hbt]
\centering
\includegraphics[width= 1.\columnwidth]{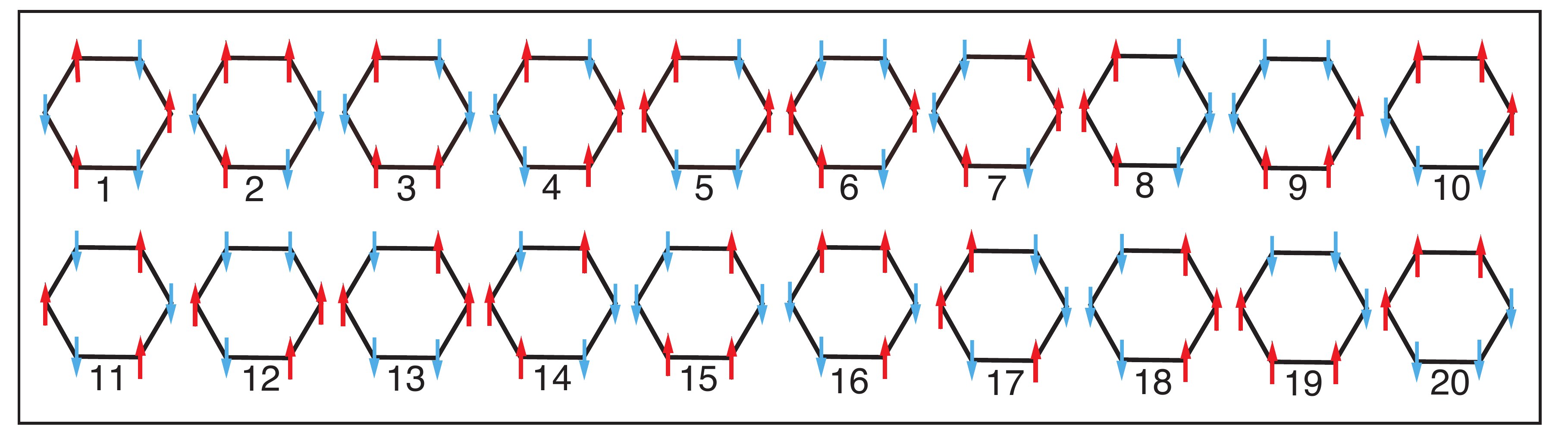}
\caption{\small{The 20 different degenerate classical ground state configurations which obeys  $\sum_{i\in\hexagon}S_z^i=0$ in a single hexagon plaquette.}}
\label{fig:3-3} 
\end{figure}

{\em Single plaquette case}: In the classical limit ($J_{\perp}=0$) of the Hamiltonian  $\hat H$ (equation \ref{ham1}) in a single plaquette, the ground state manifold is featured by 20-fold degenerate states. Each of them obeys the constraint $\sum_{i\in\hexagon}S_z^i=0$, i.e., the total magnetization around the hexagon is zero or, equivalently, three spins are "up" and the other three spins are "down". All such configurations are shown in figure \ref{fig:3-3}. The constraint is akin to the well-known ice rule "2-in 2-out" for the orientation of four dipoles arranged at the corners of a tetrahedron in a crystal ice. For the spin-version of the ice setup, so called the "spin-ice", the ice rule reads as "2-up 2-down" \cite{nisoli2013,dgf_alex_14-1} and can be mapped into a square lattice in which around each vertex four spins are arranged according to ice-rule. There are six such  possible spin-configurations. This six-vertex or ice model required that all the four spins around the vertex interact with the same strength, similar to the BFG model in which all the spins around the hexagon have the same pairwise interaction strength. In our ion-setup, there are imperfections from a perfect plaquette pattern as we have discussed above, which lift the 20-fold degeneracy partially. However, as imperfections are mostly involving spins within or on adjacent plaquettes, the investigation of a two plaquette system is required to access their effects.

\begin{figure}[hbt]
\centering
\includegraphics[width= 1.\columnwidth]{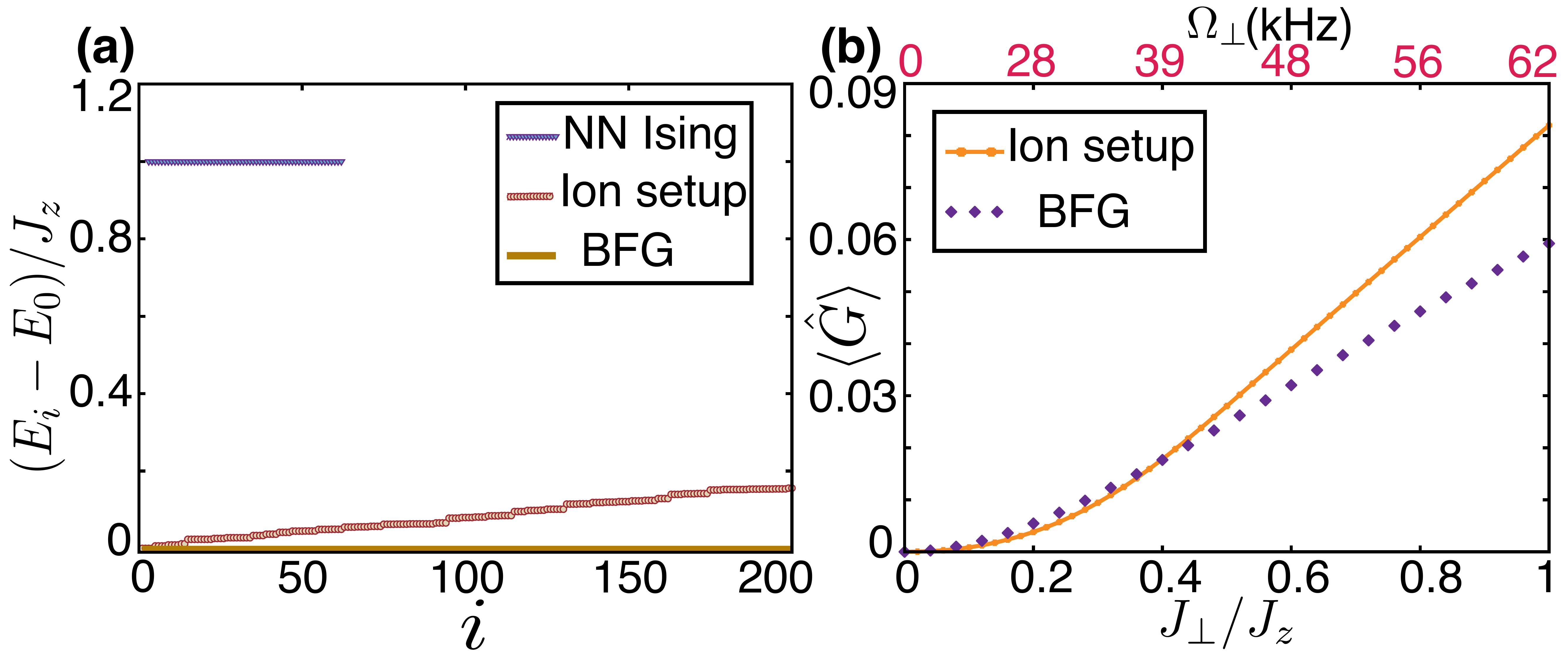}
\caption{\small{(a) The lowest 200 classical energy states in a two-plaquette system for different spin Hamiltonians with anti-ferromagnetic (AF) interactions. $E_i$ is the energy of the $i^{\rm th}$ state and $E_0$ is the ground state energy. The BFG model possesses 200 degenerate states (brown line) obeying the plaquette constraint of 3-in 3-out and in the ion setup this degeneracy is lifted by the imperfections, but the states are still very closely spaced in the spectrum. For the nearest-neighbour (NN) Ising model, there are  doubly degenerate AF ground states and the next excited states are separated by an energy of $J_z$. (b) The expectation value of $\hat G$ in the ground state as a function of the scaled hopping parameter $J_{\perp}/J_z$. In the trapped-ions quantum simulator, $J_{\perp}$ is varied by means of Rabi frequency $\Omega_{\perp}$ (see text). Due to the imperfections in the interactions for the trapped ion case, the ratio  $J_{\perp}/J_z$ is defined between their maximum values. The small values of $\langle\hat G\rangle$ even for a sufficiently large values of $J_{\perp}$ guarantees us the the stability of gauge-invariant low energy manifold. The upper axis for the Rabi frequency (corresponding to the $J_{\perp}$ in the lower axis) is estimated for the particular case when $\Omega_I=2\pi \times 500$ kHz, $\delta_m=2\pi \times 20$ kHz and $\delta_\perp=-2\pi\times 60$ kHz as discussed in section \ref{ca-ion}. In calculating $\langle\hat G\rangle$ for the BFG model, we restrict the hopping in the five spins located at the intersection of the two plaquettes, which makes sense when considering the system in the thermodynamic limit and also for the hopping pattern shown in figure \ref{fig:jhop}d for the real situation.}}
\label{fig:e_cl} 
\end{figure}

{\em Double plaquette case}: The number of degenerate ground states increases with the system size, and the degree of degeneracy is  200  for two plaquettes with the Ising part of the BFG model. In the ion setup, as in the previous case, the imperfections lift this degeneracy partially, but the energey levels are closely spaced in the energy spectrum as is evident from figure \ref{fig:e_cl}a. In our particular example for the ion setup,  the classical ground states are featured by four degenerate sates and even a very low temperature can lead to a classical order-by-disorder phenomenon in which the system collapse into one of the ground state configuration. To show the contrast, we compare the above results with a nearest-neighbour Ising model in two-plaquettes, with no inter-plaquette spin-spin interactions, the ground states are doubly degenerate (antiferromagnetic states) and any excited state, resulting from a single spin flip, leads to an energy cost of $J_z$, see figure \ref{fig:e_cl}a. This implies that our ion setup for the implementation of plaquette interactions is indeed a promising candidate for emulating frustrated quantum magnetism and en route to quantum spin liquid once scaled up to larger systems.

In the presence of quantum dynamics the global magnetization, $M=\langle\sum_i\hat S_z^i\rangle$  is preserved, but the magnetization per palquette is not.  We define the following operator to quantify the magnetization per plaquette,
\begin{equation}
\hat G=\frac{1}{N_{\hexagon}}\sum_{\hexagon}\left(\sum_{i\in \hexagon} S_z^i\right)^2,
\end{equation}
where $N_{\hexagon}$ is the number of hexagonal plaquettes ($=2$ in our case). The magnitude of $\langle G \rangle$, the expectation value taken in the ground state of the spin-Hamiltonian, gives us the measure of the admixture from states outside the classical ground state manifold obeying the plaquette constraint. In practice, this determines the quality of gauge-invariance as a function of quantum fluctuations: indeed, $\hat{G}$ can be viewed as an effective generator of a local (gauge) symmetry, which commutes with the effective Hamiltonian in the strong coupling $J_z\gg J_\perp$ limit~\cite{sheng2005}.

With no hopping ($J_{\perp}=0$), $\hat{G}=0$, and it increases as $J_{\perp}$ increases as shown in figure \ref{fig:e_cl}b. The small values of $\langle\hat G\rangle$ even for sufficiently large values of $J_{\perp}$ shows the stability of the low energy manifold with a plaquette constraint against quantum fluctuations. In both cases, as expected from perturbation theory arguments, $\langle \hat G\rangle \sim (J_{\perp}/J_z)^2$ close to the classical limit. We note that the only effect of the imperfections is quantitative: some {\em gauge variant} states not satisfying the plaquette constraint pay a smaller energy penalty compare to the ideal BFG case. However, their effect is relatively small even for intermediate coupling strengths.

\subsection{Plaquette-magnetization dynamics: deconfined charge excitation}
\begin{figure}[hbt]
\centering
\includegraphics[width= 1.\columnwidth]{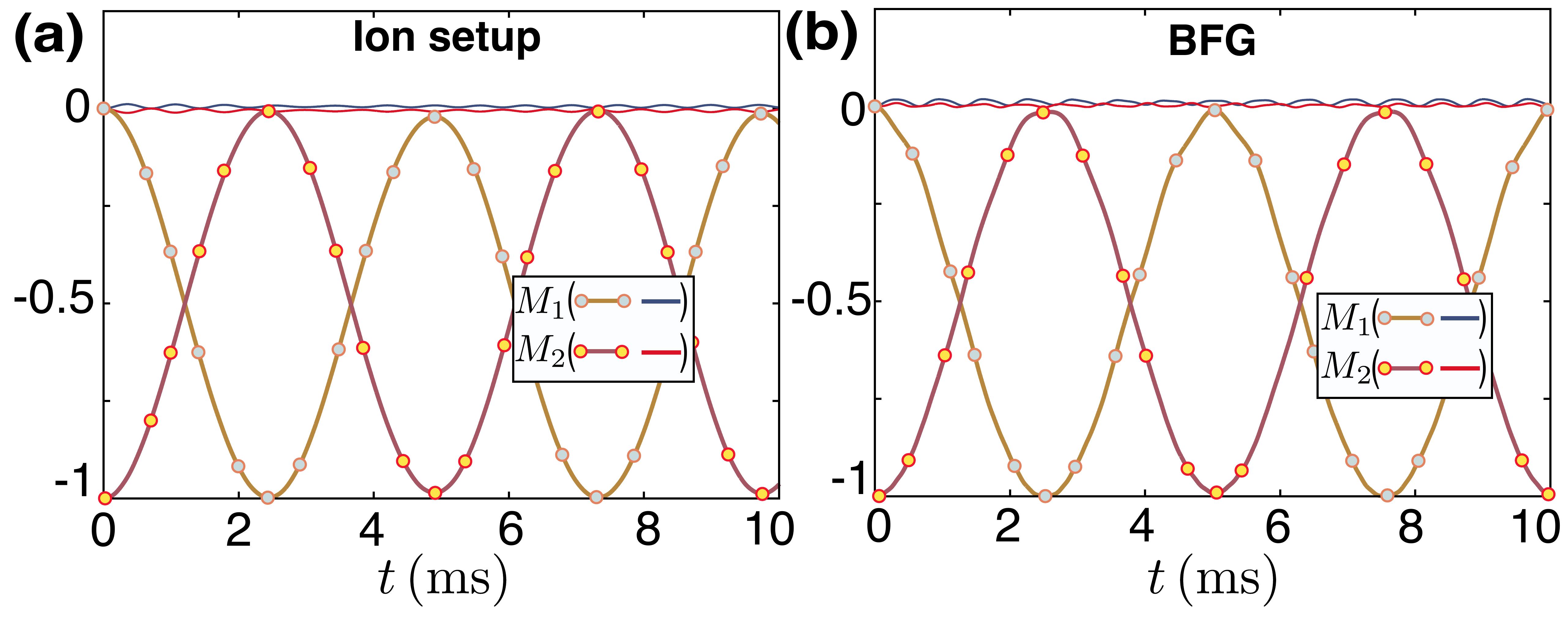}
\caption{\small{The time propagation of plaquette magnetizations $M_1$ and $M_2$ for (a) the BFG model and (b) the ion setup with $J_{\perp}/J_z=0.2$ (for the ion setup the ratio is taken with the maximum values of $J_z$ and $J_{\perp}$). The solid lines are for the case in which the initial state is prepared in a state in which the $M_1=M_2=0$ and the magnetization is hardly affected by the hopping dynamics as expected. The lines with bubbles are the case for which the initial state is such that one of the plaquette carries a charge (in this particular example a negative charge in the second plaquette, i.e., $M_2=-1$) and the other plaquette has zero charge (or zero magnetization, $M_1=0$). In time the charge oscillates between the two plaquettes indicating a de-confined phase.}}
\label{fig:mdyn} 
\end{figure}

In this section we examine and compare the dynamics of magnetization in each plaquette for an ion setup with plaquette interactions with the BFG model, for different initially prepared spin configurations. The local plaquette magnetizations are defined as $M_1=\sum_{i\in\hexagon^1}S_z^i$ and $M_2=\sum_{i\in\hexagon^2}S_z^i$ respectively for plaquettes 1 and 2. To address the stability of gauge constraint against spin hopping, we study the time propagation starting from one of the classical degenerate ground state, which obeys  3-in 3-out rule. As expected, the hopping dynamics hardly affected the local plaquette magnetizations even for large hopping parameter $J_{\perp}$. A particular example with $J_{\perp}/J_z=0.2$ is shown in figures \ref{fig:mdyn}a and \ref{fig:mdyn}b for the ion setup and the BFG model, respectively, and they are found to exhibit the same behaviour. The dynamics become more interesting when we introduce a {\it charge} in one of the plaquttes by flipping a single spin in the respective plaquette. As an example we consider an initial state in which $M_1=0$ and $M_2=-1$ and the inter-plaquette spin hopping couples this state to $M_1=-1$ and $M_2=0$ and hence the state oscillates between them. This has a clear interpretation in gauge theoretical language: the "doped" plaquette represents a charge moving on top of a gauge-invariant background. This scenario can be compared to a case where a pair of quasi-particle and quasi-hole (fractionalized charges or excitations) is created by a single spin flip in a square lattice, which supports an exact incompressible quantum liquid. The particle-hole pair doesn't form a bound pair but is in a deconfined phase \cite{chern_13}. The same is also predicted for a classical Ising model in pyrochlore lattice in which the macroscopic ground state degeneracy leads to the deconfinement of monopole excitations \cite{mono_moess_07}. This is one of the intriguing point of this paper from a frustrated magnetism picture, arising with a question on the stability of this deconfined phase, interestingly, which could be qualitatively addressed with a small size ion-crystal quantum simulator as we show here.

The many-body physics associated with the reduced Hamiltonian $H_{XZ}$ mentioned in section \ref{model} is discussed in \ref{hxy} and found to possesses the same properties as that of the full Hamiltonian.

\section{Implementation with a 2D crystal of $^{40}$Ca$^+$}
\label{ca-ion}

As an example, we consider a crystal of $^{40}$Ca$^+$ ions. Since this ion does not posses a nuclear spin, we employ the Zeeman and metastable electronic states, rather than  hyperfine ground states. We encode the spins in electronic states of the ion, $|\downarrow\rangle = |S_{1/2},m_j=1/2\rangle$ and $|\uparrow\rangle = |D_{5/2},m_j=3/2\rangle$~\cite{Benhelm_NP_08}. Inititialization can be performed by optical pumping followed by addressed coherent laser pulses that prepare each ion in a designated spin state. In this way the ions that will be pinned can also be prepared in the state  $|1\rangle = |S_{1/2},m_j=-1/2\rangle$. The pinning can be done using e.g. a $\lambda_{pin}$~=~532~nm retro-reflected laser with a power of $\sim$~50~mW, focussed to a waist of~2~$\mu$m. For an estimate, in which we are neglecting the polarizabilities of all transitions except the dominant $S \rightarrow P$ transition around 397~nm, this results in an anti-trapping potential of trap-frequencies (0.08,0.08,1.3)~2$\pi$ MHz in a node of the standing wave. With a Paul trap of transverse frequency $\omega_{\perp}=2\pi$~3~MHz, this results in a  local reduction of the trap-frequency of $\sim$~10~\% to $2\pi$~2.7~MHz as in the example in section~\ref{sec_single_p}. The trap-frequency in the other two directions is hardly affected (it is reduced by $\sim 2\pi$~3~kHz). 

Errors in the realization of the BFG model arise from spontaneous scattering from the pinning laser, wich is reduced because of the large detuning and because the ion is placed in a node of a wave. We expect a scattering rate of $\leq$~1~s$^{-1}$ as an upper bound. The ions are separated by about 5~$\mu$m, which is much larger than the waist of the pinning laser, such that the ions in the plaquette are not affected by it. 

From an experimental point of view, it is convenient to perform a change of basis to exchange $\sigma_z$ and $\sigma_x$, as spin-spin interactions in the latter require less laser power~\cite{Roos_NJP_08}, whereas we are interested in the regime $J_z>J_{\perp}$. Spin-spin interaction terms involving $\sigma_y$ can be ommitted in a first implementation and this results in non-resonant $S^+ \otimes S^+$ and $S^-\otimes S^-$ terms that are suppressed by the Gauss law, shown in \ref{hxy}. The $J_z$ plaquette spin-spin interaction can now be induced by a M{\o}lmer-S{\o}rensen scheme~\cite{qc_mol_99}, operating on the optical qubit at 729~nm~\cite{Benhelm_NP_08,Kirchmair_NJP_09}. In particular, a bichromatic light field with components around $\omega_{\uparrow\downarrow}\pm \omega_m \pm \delta_m$ induces $S_x \otimes S_x$ interactions~\cite{Roos_NJP_08}. Rabi frequencies of $\Omega_I=2\pi$~500~kHz and a detuning $\delta_m=2\pi$~20~kHz result in (plaquette) spin-spin interactions of $J_z\eqsim2\pi$~18.5~kHz. The pinned ion would have a transient maximal amplitude of oscillation of around 20~nm~$\ll \lambda_{pin}$, such that the harmonic approximation is justified for the pinning laser. The fidelity of the gate for $\sigma_z$ is calculated semi-classically and its lower bound is about 99.7$\%$, limited by remaining spin-motion entanglement in far detuned phonon modes. For details regarding the fidelity calculation see \ref{fid}. To realise $J_{\perp}/J_z \eqsim 0.2$, we require a Rabi frequency for the hopping spin-spin interactions of about $\Omega_I\eqsim 2\pi ~ 20$ kHz with $\delta_{\perp}=-2\pi ~  60$ kHz as discussed above. 

For ions with a nuclear spin, such as $^{9}$Be$^+$or $^{171}$Yb$^+$, the three required long-lived states can be encoded in the hyperfine ground states and the spin-spin interactions can be induced by Raman lasers, which can reach comparable coupling strengths as for the electronic state encoding descibed above~\cite{rvw_wine_03}. For Yb$^+$ strong optical pinning has been shown recently~\cite{ol_ion_vule_13}. 

The readout of the quantum simulation run is performed by detecting state-dependent laser induced resonance fluorescence, where the state $|\downarrow\rangle$ will appear bright while sites with ions in $|\uparrow\rangle$  remain dark as the state does not scatter photons. Detection fidelities exceed 99\% in ion trap experiments routinely.  Note that the detection can easily resolve individual ions, such that the outcome can be immaged on a ccd chip in one picture, where populations, and correlations may be readily revealed.  This allows not only for obtaining the global magnetization but also to determine correlations of spins such that the gauge-invariant dynamics can be observed in the plaquettes.

\section{Conclusion and Outlook}
\label{summ}

In conclusion, we have shown that using phonon mode shaping it is possible to simulate exotic spin-spin interactions in an ion crystal which have far reaching consequences in the context of strongly correlated systems, in particular frustrated magnetic systems in which low energy manifold is described by an emergent dynamical gauge field. The mode-shaping is accomplished by (anti-) pinning standing wave light fields, which modify locally the ion trap and affect its motional dynamics.  Alternatively, we can selectively excite ions to a Rydberg state ~\cite{Muller_NJP_09,FSK_NJP_11} in which those ions experience a different trapping potential compared to the ground state ions due to state-dependent atomic polarizabilities, see \ref{rion}.  Additional microwave fields could be employed for fine-tuning~\cite{Li_PRL_12,Li_PRA_13} the trapping field of the ions occupying the Rydberg state. This leads to a design of the tranversal mode structure in the planar crystal, similar to that with optical pinning forces. 

From a quantum magnetism point of view, we have discussed the ground state and magnetization dynamics of the corresponding spin Hamiltonian of the proposed trapped ion implementation and the results have been compared with that of the idealized BFG model, within a small setup of one and two plaquettes. The results show the excellent agreement between the physics  in an ion based quantum simulator and that of the BFG model. The deconfined dynamics of a single charge excitation in a two plaquette system, arising from the ring exchange, is akin to the deconfined monopole excitations in a 3D pyrochlore spin-ice, arising from the large macroscopic ground state degeneracy. In general, the current studies would open up a completely new aspect of studying spin-Hamiltonians using ultra cold ion crystals, in particular once scaled up, will be the prime quantum simulator for emulating topological quantum spin liquids or resonating-valence-bond  states.

\section*{Acknowledgements}

We thank R. Moessner and M. M\"uller for interesting discussions. We acknowledge funding through the ERA-NET within the Ryd-ION consortium. FSK and RG acknowledge funding by the EU STREP EQuaM, SIQS and by the DFG via Sonderforschungsbereich SFB-TR/49. Work at Innsbruck is supported by ERC Synergy Grant UQUAM, COHERENCE, SIQS, and SFB FoQuS (FWF Project No. F4006).

\appendix

\addcontentsline{toc}{section}{Appendices}
\addtocontents{toc}{\protect\setcounter{tocdepth}{-1}}
\section{Atom-light interactions and $\sigma_z\otimes \sigma_z$ interactions}
\label{ss-int}
In this section we discuss briefly the interaction of Raman fields with the spin states $\{|\downarrow\rangle, |\uparrow\rangle\}$ and the resulting spin-spin interactions between the ions. The Hamiltonian for N harmonically trapped two level ions interacting with a laser field of frequency $\omega_I$ is given by
\begin{equation}
\hat H=\hat H_0+\hat H_I,
\end{equation}
where
\begin{equation}
\hat H_0=\frac{\hbar\omega_{\downarrow\uparrow}}{2}\sum_{i=1}^N\sigma_z^i+\sum_{m=1}^{m=3N}\hbar\nu_m\hat a^{\dagger}_m\hat a_m,
\label{h01}
\end{equation}
where $\sigma_z$ is the Pauli spin-1/2 matrix and the operator $\hat a_m$ $(\hat a_m^{\dagger})$ annihilates (creates) a phonon in the $m^{th}$ mode with an eigen value $\nu_m$ (see \ref{eqbm}).
In the interaction picture, the atom-light interaction Hamiltonian \cite{schn-rvw12, lieb-rvw03, wine-rvw98} reads as,
\begin{eqnarray}
\hat H_I=\sum_i\hbar\Omega_I^i \left[e^{i\sum_{m=1}^{m=N}\eta_m^i\left(\hat a_me^{-i\nu_mt}+\hat a^{\dagger}_me^{i\nu_mt}\right)-i\omega_It+\phi_I^i}+{\rm h.c.}\right]\hat\kappa^i.
\label{H-int}
\end{eqnarray}
where $\Omega_I^i=-d \mathcal E^i/2 \in \mathbb R$ is the effective coupling strength for $i^{th}$ ion with $d$, the dipole moment for the transition and $\mathcal E^i$ is the electric field strength at the $i^{th}$ ion position, $\eta_m^i=q_{m0}b_m^ik_I$ are the Lamb-Dicke parameters with $q_{m0}=\sqrt{\hbar/2M\nu_m}$ and ${\bf b}_m=\{b_m^i\}$ is the phonon eigen vector, and $\phi_I$'s are additional phases in each ion.  The operator $\hat {\kappa}^i$, a general $2\times 2$ matrix, is determined by the polarization of the electric field and the atomic states. Note that the summation of $m$ in the exponential term in equation \ref{H-int} is only over the $N$ transversal $z$-modes and we replaced the position operators of the ions in terms of normal mode operators while writing the Hamiltonian $H_I$ \cite{schn-rvw12}.

In the weak coupling limit: $\Omega_I\ll\omega_{\downarrow\uparrow}$, and when $\omega_I\ll\omega_{\downarrow\uparrow}$ (an implementation in this limit is shown in figure \ref{fig:setup}b) and in the Lamb-Dicke regime: $\eta_m^i\sqrt{\langle(\hat a_m+\hat a_m^{\dagger})^2\rangle} \ll 1$ where we keep only the terms upto first order in the Lamb-Dicke parameters $\eta_m^i$, then taking the rotating wave approximation (RWA) in which the terms with $e^{\pm i\omega_{\downarrow\uparrow}t}$ are neglected, we finally get the atom-light interaction Hamiltonian as
 \begin{eqnarray}
\hat H_I^z=\sum_i\hbar\Omega_I^i \sum_{m=1}^{N}\eta_m^i\hat a^{\dagger}_me^{-i\delta_mt+i\phi_I} \sigma_z^i+{\rm h.c.},
\label{H-I-z}
\end{eqnarray}
where $\delta_m=\omega_I-\nu_m$. From the form of equation \ref{H-I-z} it is clear that there is no spin flip involved which is justified by the limit $\omega_I\ll\omega_{\downarrow\uparrow}$. In general, equation \ref{H-I-z} describes quantum gates based on an effective laser field being interacting with many ions. The electric field from the laser beams gives rise to a Stark shift for each internal states, and a corresponding electric-dipole force on each ion. These state dependent forces are typically used in trapped-ion quantum computation \cite{qc_mol_99,ga_zoll_95}. The time evolution of the system is then obtained using the Magnus expansion \cite{mag-exp54}  for the time evolution operator
\begin{equation} 
\hat U(t,0)=\exp\left[\frac{-i}{\hbar}\int_0^tdt^{\prime} H(t^\prime)-\frac{1}{2\hbar^2}\int_0^tdt^{\prime}\int_0^{t^{\prime\prime}}[H(t^{\prime},H(t^{\prime\prime}))]\right],
\end{equation}
where $H(t)$ is the time-dependent Hamiltonian. In our case, with sufficiently large detuning from the motional sideband \cite{kim_tm_09} and assuming the same phase for all ions, the long-term time evolution is dominated by a term linear in time, $t$ and is of the form $\exp(-i\sum_{i,j}J_z^{ij}\hat\sigma_z^i\hat\sigma_z^jt/4)$ \cite{mg_qs_08} with 
\begin{equation}
J_z^{ij}=\sum_{m=1}^{N}\frac{4\Omega_I^i\Omega_I^j  \eta_m^i\eta_m^j}{\delta_m}.
\end{equation}
 This realizes an Ising spin-Hamiltonian of the form $\hat H_z=\sum_{i,j}J_z^{ij}S_z^i\otimes S_z^j$ with spin-spin couplings $J_z^{ij}$ between $i^{th}$ and $j^{th}$ ions. Note that the spin-$1/2$ operators  ($S_{\alpha}$) defined in equation \ref{ham1} are related to the Pauli spin-$1/2$ matrices ($\sigma_{\alpha}$)  through $S_{\alpha}=\sigma_{\alpha}/2$.
 

\section{Ion crystal: Equilibrium positions and vibrational spectrum}
\label{eqbm}
In a 3D setup, the potential experienced by the ions is 
 \begin{eqnarray}
 V=\sum_{i=1}^{N}\frac{1}{2}M\left(\omega_x^2x_i^2+\omega_y^2y_i^2+\omega_z^2z_i^2\right)+
  \sum_{i,j=1, i\neq j}^{N}\frac{Ze^2}{8\pi\epsilon_0}\frac{1}{|{\bf r}_i-{\bf r}_j|},
\end{eqnarray}  
where $|{\bf r}_i-{\bf r}_j|=\sqrt{(x_i-x_j)^2+(y_i-y_j)^2+(z_i-z_j)^2}$. The first term is the harmonic confinement due to external fields and the second term accounts for the Coulomb interaction between the positively charged ions. We choose the length scale as $l_x^3=(Z^2e^2)/(4\pi\epsilon_0 M\omega_x^2)$ and in the dimensionless form the potential reads as
 \begin{equation}
 V=\sum_{i=1}^{N}\frac{1}{2l_x^2}\left(x_i^2+\lambda_y^2y_i^2+\lambda_z^2z_i^2\right)+\frac{1}{2l_x}\sum_{i,j=1, i\neq j}^{N}\frac{1}{|{\bf r}_i-{\bf r}_j|},
\end{equation}  
where $\lambda_z=\omega_z/\omega_x$, $\lambda_y=\omega_y/\omega_x$ . In the following we absorb $l_x^{-1}$ in the position co-ordinates such that $x_i$, $y_i$ and $z_i$ are dimensionless. 
 The equilibrium positions are calculated by solving the equations 
\begin{equation}
\frac{\partial V}{\partial x_i}=\frac{\partial V}{\partial y_i}=\frac{\partial V}{\partial z_i}=0,
\end{equation}
for each ions. This leads to solving the coupled algebraic equations of the form
\begin{equation}
\lambda_{\alpha}^2\alpha_{m}+\sum_{i=1, i\neq m}^N\frac{(\alpha_i-\alpha_m)}{[R_0(i,j)]^{3/2}}=0,
\label{eqbmpns}
\end{equation}
where $\alpha=\{x,y,z\}$ and $R_0(i,j)=(x_i-x_j)^2+(y_i-y_j)^2+(z_i-z_j)^2$. Once the equilibrium positions are obtained, we can calculate the eigen vectors ${\bf b_m}$ and eigen values $\nu_m$ of the phonon modes by the exact diagonalization of the Hessian matrix constructed out of the following second order derivatives:
\begin{itemize}
\item {for $m=n$}
\begin{equation}
\frac{\partial^2V}{\partial \alpha_n\partial \alpha_m}=\lambda_{\alpha}^2+\sum_{i\neq m}^{N}\left[\frac{3(\alpha_i-\alpha_m)^2}{R_0(i,m)^{5/2}}-\frac{1}{R_0(i,m)^{3/2}}\right]
\end{equation}
\begin{equation}
\frac{\partial^2V}{\partial \alpha_n\partial \beta_m}=\sum_{i\neq m}^{N}\left[\frac{3(\alpha_i-\alpha_m)(\beta_i-\beta_m)}{R_0(i,m)^{5/2}}\right],
\end{equation}

\item{for $m\neq n$}
\begin{equation}
\frac{\partial^2V}{\partial \alpha_n\partial \alpha_m}=-\frac{3(\alpha_n-\alpha_m)^2}{R_0(n,m)^{5/2}}+\frac{1}{R_0(n,m)^{3/2}}
\end{equation}
\begin{equation}
\frac{\partial^2V}{\partial \alpha_n\partial \beta_m}=-\left[\frac{3(\alpha_n-\alpha_m)(\beta_n-\beta_m)}{R_0(n,m)^{5/2}}\right],
\end{equation}
\end{itemize}
with $\alpha,\beta\in \{x,y,z\}$ and $\alpha\neq\beta$. 

\section{Semi-classical estimation of population in the motional states and gate fidelity}
\label{fid}
The phonon spectrum,
\begin{equation}
H_m=\hbar\nu_m\sum_m\hat a^{\dagger}_m\hat a_m
\end{equation}
is described as a collection of harmonic oscillators with frequencies $\nu_m$. As described in the text, the quantum emulation of the BFG Hamiltonian relies on the phonon-mode shaping by optically anti-pinning few ions in the crystal. Our focus will be on low lying transverse modes, which gives us a hexagonal plaquette pattern when coupled dressed by Raman laser fields. Any discrepancy  from the exact BFG model in the emulation results from the population of other modes which is not of our prime interest.  We semi-classically investigate the population of these higher modes, and provide an estimation for the fidelity of the corresponding gate operations. The problem is equivalent to an undamped, driven harmonic oscillator in which the driving force is a sinusoidal one with frequency $\omega_I$, that of the laser field. The governing equation is 
\begin{equation}
\frac{d^2z}{dt^2}+\nu_m^2z=\frac{\hbar\Omega_Ik_0}{M}\cos(\omega_It),
\label{fho}
\end{equation}
where $k_0=\eta_m/q_{m0}$ with $q_{m0}=\sqrt{\frac{\hbar}{2M\nu_m}}$. For the initial conditions $z(0)=0$ and $z^{\prime}(0)=0$, we get the solution
\begin{equation}
z(t)=\frac{\hbar\Omega_Ik_0}{M\left(\nu_m^2-\omega_I^2\right)}\left[\cos \omega_It-\cos \nu_mt\right].
\end{equation}

\begin{figure}[hbt]
\vspace{0.5cm}
\centering
\includegraphics[width= .5\columnwidth]{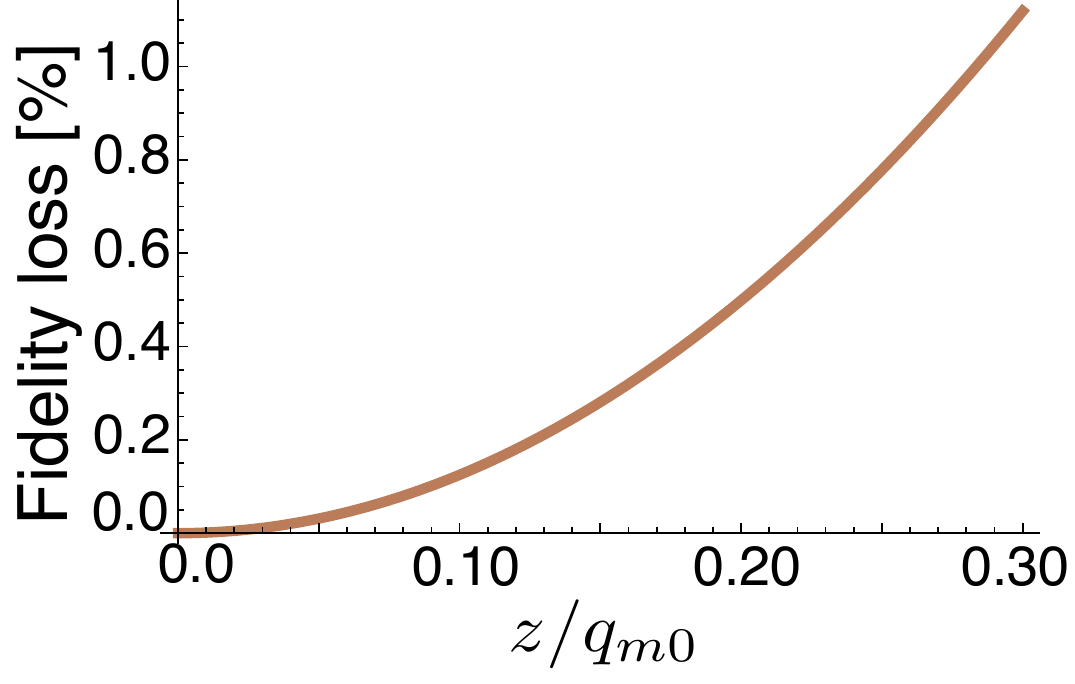}
\caption{\small{The percentage of fidelity loss as a function of the displacement $z/q_{m0}$. To keep the fidelity loss with in $1\%$, the displacement  $z/q_{m0}$ required to be roughly $\leq$0.25.}}
\label{fig:floss} 
\end{figure}

The classical, (maximum) amplitude of oscillation, $A\simeq\frac{2\hbar\Omega_Ik_0}{M\left(\nu_m^2-\omega_I^2\right)}$ can be related to the quantum mechanical spectrum of Harmonical oscillator by equating the classical and quantum energies, and we get
\begin{equation}
n_m=\frac{1}{4}\left(\frac{A}{q_{m0}}\right)^2,
\label{n-ph-m}
\end{equation}
after we left out the zero point oscillation energy. This relation would provide us a rough estimation for the number of phonons in the system for a particular mode $m$. In addition, the overlap between the initial phonon state and the maximally displaced state would give us the fidelity loss arising from this particular mode. We require this loss would be well below $1\%$. Taking the Gaussian ground state solution of the harmonic oscillator, we get the overlap as a function of the displacement $z_0$ as 
\begin{equation}
\mathcal O_m=\exp(-z_0^2/8q_{m0}^2).
\label{ovlap}
\end{equation}
The fidelity loss $\mathcal L_m=1-\mathcal O_m$ due to a given mode $m$, as a function of the amplitude of oscillation is shown in Fig. 
\ref{fig:floss}, and one can see that the larger the displacements, the lesser the fidelity is. In order to keep the fidelity loss with in $1\%$, the displacement  $z/q_{m0}$ required to be $\leq$0.30. The net fidelity loss is then obtained by the summing over that of all modes.
\section{Gauss's law and plaquette magnetization dynamics for the reduced Hamiltonian, $H_{XZ}$}
\label{hxy}
 \begin{figure}[hbt]
\centering
\includegraphics[width= 1.\columnwidth]{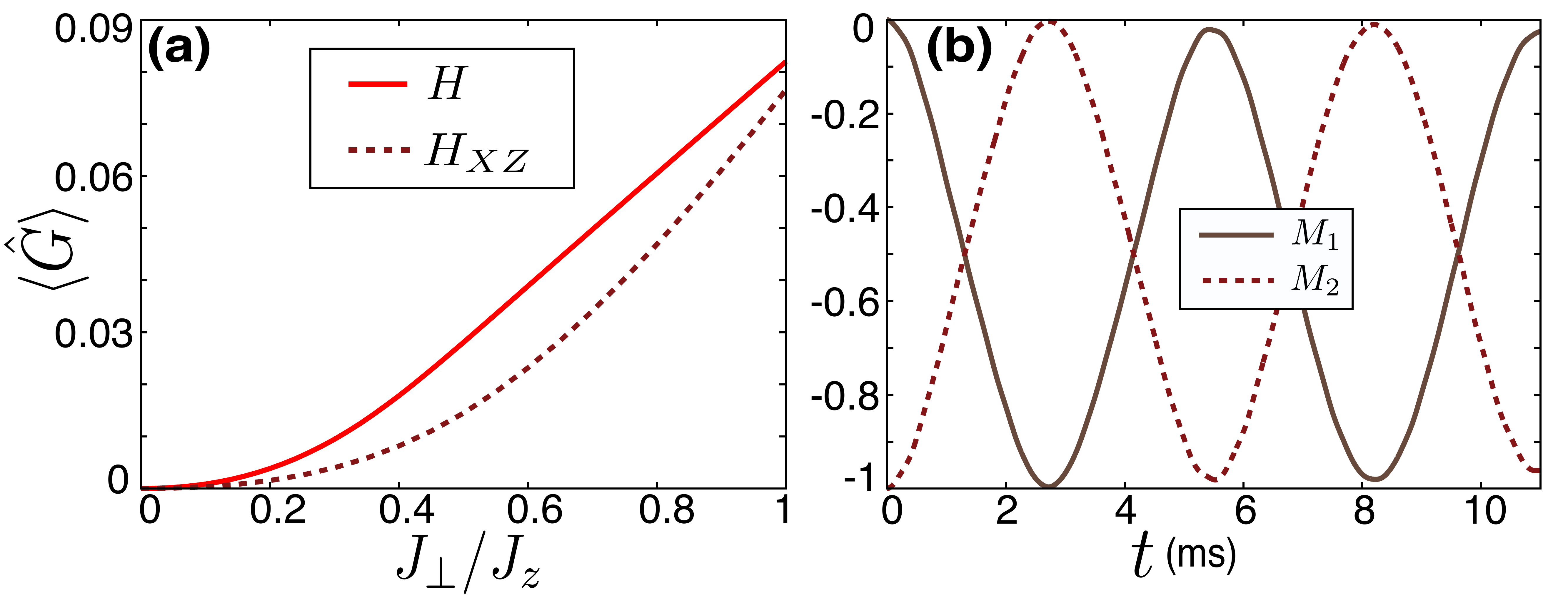}
\caption{\small{(a) Gauss's law for the reduced Hamiltonian $H_{XZ}$ compared to the full Hamiltonian $H$ in an ion-setup for the same parameters as in figure \ref{fig:e_cl}. (b)The time propagation of plaquette magnetizations $M_1$ and $M_2$ for the reduced Hamiltonian with $J_{\perp}/J_z=0.4$ (the ratio is taken with the maximum values of $J_z$ and $J_{\perp}$). The initial state is prepared such that $M_1=0$ and $M_2=-1$.}}
\label{fig:gl2} 
\end{figure}

In this section we consider a reduced Hamiltonian of the form
\begin{equation}
H_{XZ}=\sum_{i<j}J_z^{ij}S_z^i\otimes S_z^j+\sum_{i<j}J_{\perp}^{ij}S_x^i\otimes S_x^j.
\end{equation}
Experimentally, it is advantageous to have the reduced Hamiltonian over the full one since less number of independent lasers are involved. The coupling parameters  $J_z^{ij}$ and $J_{\perp}^{ij}$ are still the same which we discussed in the main part of the text. 
Contrary to the full Hamiltonian, here it involves not only {\em flip-flop} terms but also {\em flip-flip} and {\em flop-flop}, as seen from the expansion, $S_xS_x=(S_+S_-+S_-S_++S_-S_-+S_+S_+)/4$. The results are shown in figure \ref{fig:gl2} and the resulting Gauss's law is compared with that of  the full Hamiltonian $\hat H$. The hopping of charge excitation is also demonstrated, revealing the high flexibility of trapped-ion quantum simulators. 

\section{Rydberg excitations in an ion crystal: atomic polarizibility and modified trap frequency}
\label{rion}
Instead of using the pinning lattices on ions in state $|1\rangle$ (see main text), we can as well excite or weakly dress those ions to a Rydberg state. Due to the state dependent atomic polarizabilities, the Rydberg excited ions experience different trapping frequencies compared to those occupying the low-lying atomic states. This modifies the phonon spectrum of the ion crystal, identical to the situation in which the pinning lattices are present. Recently, it has been shown \cite{Li_PRL_12, Li_PRA_13} that ions excited to the $nP$-Rydberg state experience an additional radial potential, $V_a({\bf r})\approx -e^2\alpha^2\mathcal{P}_{nP}r^2$, where the polarisability $\mathcal{P}_{nP}\approx-0.25\times n^7$ atomic units, with $n$, the principal quantum number of the Rydberg state. Effectively, an ion in this Rydberg state experiences a tighter confinement compared to an ion occupying a low-lying state, but with additional microwave fields, by coupling to a nearby Rydberg state, say $n^{\prime}S$, 
we can freely tune the trapping frequency of the Rydberg excited ion. It can be increased, decreased or even made equal to that of ions occupying the qubit states $|\uparrow\rangle$ and $|\downarrow\rangle$. The dressed states  $|\pm\rangle=N_{\pm}(C_{\pm}|nP\rangle+|n'S\rangle)$ polarizabilities read as

\begin{equation}
\mathcal{P}_{\pm}=N_{\pm}^2\left(C^2_{\pm}\mathcal{P}_{np}+\mathcal{P}_{n^{\prime}S}\right)
\end{equation}
with $\mathcal{P}_{n^{\prime}S}>0$, the parameter $C_{\pm}$ depends on the microwave parameters such as detunings and Rabi frequencies and $N_{\pm}$ are the normalization constants. More details on the microwave-Rydberg approach can be found in \cite{Li_PRA_13}. $\mathcal{P}_{\pm}$ are easily controlled via microwave field paremeters, and hence the local trapping of Rydberg ions. The only limit in this approach is set by the life time of Rydberg states for gate operations and it can be augmented by going either to a higher Rydberg level or by weakly dressing to the Rydberg state.

\section*{References}

\providecommand{\newblock}{}
\bibliography{liball.bib}

\end{document}